# STUDENTS' PERCEPTIONS OF SUSTAINABLE UNIVERSITIES IN HUNGARY: AN IMPORTANCE-PERFORMANCE ANALYSIS

**Szabolcs Nagy[1*] and Mariann Veresné Somosi[2]**
[1)2)] *University of Miskolc, Miskolc, Hungary*

**Abstract**
In order to succeed, universities are forced to respond to the new challenges in the rapidly changing world. The recently emerging fourth-generation universities should meet sustainability objectives to better serve their students and their communities. It is essential for universities to measure their sustainability performance to capitalise on their core strengths and to overcome their weaknesses. In line with the stakeholder theory, the objective of this study was to investigate students' perceptions of university sustainability including their expectations about and satisfaction with the efforts that universities make towards sustainability. This paper proposes a new approach that combines the sustainable university scale, developed by the authors, with the importance-performance analysis to identify key areas of university sustainability. To collect data, an online survey was conducted in Hungary in 2019. The sustainable university scale was found to be a reliable construct to measure different aspects of university sustainability. Results of the importance-performance analysis suggest that students consider Hungarian universities unsustainable. Research findings indicate that Hungarian universities perform poorly in sustainable purchasing and renewable energy use, but their location and their efforts towards separate waste collection are their major competitive advantages. The main domains of university sustainability were also discussed. This study provides university decision-makers and researchers with insightful results supporting the transformation of traditional universities into sustainable, fourth-generation higher education institutions.

* Corresponding author, **Szabolcs Nagy** – nagy.szabolcs@uni-miskolc.hu

**Introduction**

We live in the age of rapid changes to which higher education institutions should adopt. A university that wants to succeed needs to respond to the challenges of the new era. One of them is the urgency to meet sustainability objectives (Filho, Manolas and Pace, 2015; Soini, et al., 2018; Olalla and Merino, 2019). Universities are undergoing a rapid transformation as they are not only traditionally engaged in education but are also playing an increasingly important role in the society (Papp-Váry and Lukács, 2019). Nowadays, the emergence of the so-called Fourth Generation universities, which actively shape their socio-economic environment, can be seen (Pawłowski, 2009; Lukovics and Zuti, 2017).

The topic of sustainable development is increasingly present among the major concerns of the international academic community (Grecu and Ipiña, 2014). Universities must take steps to achieve the United Nations Sustainable Development Goals (Paletta, et al., 2019). Target 4.7 declares that students have the right to acquire the knowledge and skills needed to promote sustainable development (UN, 2019). Globally, the proliferation of the efforts to assess universities' responses to the challenges of sustainability can be seen (Li, Gu and Liu, 2018). Adams, Martin, and Boom (2018) draw the attention to the importance of the university sustainability culture.

Adaptation of the stakeholder theory is essential for higher education institutions (Mainardes, et al., 2010) as stakeholders can create opportunities for or pose threats to an organisation (Chapleo and Sims, 2017). Students as stakeholders have a serious impact on the future development of universities (Degtjarjova, Lapina and Freidenfelds, 2018). Commitment to sustainability of leaders and important stakeholders play a key role in the effectiveness of sustainable development initiatives in higher education institutions (Wright, 2010.)

The position of Hungarian higher education institutes in the world rankings is not very favourable. The best Hungarian university can be found around the 500th place in global rankings. There are only seven or eight Hungarian institutions that are ranked at all (Polónyi and Kozma, 2019). The weak performance of the Hungarian higher education institutions in sustainability rankings explains the need for a comprehensive analysis of university sustainability in Hungary from the students as stakeholders' perspective, which is one of the main objectives of this study.

Students as stakeholders form expectations regarding university sustainability not only generally, but also very specifically, and how those expectations are met determines the level of their satisfaction. This study aims to investigate student expectations about and satisfaction with the attributes of the sustainable university by using the sustainable university scale (SUS) combined with the importance-performance analysis (IPA). SUS, the items of which are the determinants of university sustainability, was developed by the authors. IPA has been widely used to examine the relationship between importance, performance, and satisfaction in many areas (Yuvinatileng, et al., 2013; Wyród-Wróbel and Biesok, 2017, Kim, et al., 2018) However, no previous study has investigated it in the context of university sustainability in spite of the fact that universities should use managerial tools to develop their sustainability strategy. This study seeks to address this research gap.

# 1. Literature review

## 1.1. Perceptions of the sustainable university

In the UI GreenMetric World University Ranking 2019, which provides information about the current conditions and policies related to Green Campus and Sustainability, only seven Hungarian universities can be found. The University of Szeged is in the best position, ranked first in Hungary, and 74th in the world. It is followed by the University of Pecs, ranked 100th globally and the University of Debrecen, in the 202nd position in the world ranking. The University of Miskolc, for which the authors work, can be found only in the 605th place in this ranking of 780 universities globally (Greenmetric, 2019). Students' perceptions of university sustainability were assumed to be in line with this poor ranking performance. *It is therefore hypothesized that students are not satisfied with the sustainability performance of the Hungarian higher education institutions (H1).* Mention must be made of some of the shortcomings of the GreenMetric Ranking, i.e. non-compliance with the Berlin Principles (Ragazzi and Ghidini, 2017), however, it is still one of the best tools to quantify university sustainability.

The perceptions of university students towards factors of a sustainable university was first discussed by Nejati and Nejati (2013). The authors developed a reliable scale to assess the university practices towards sustainability. They identified four main dimensions of the sustainable university scale, which are respectively: 1) community outreach, 2) sustainability commitment and monitoring, 3) waste and energy, and 4) land use and planning. Their initial scale contained 28 items, which they reduced to a 12-item scale, which could be a key instrument for university decision-makers and stakeholders to measure the university's performance regarding the implementation of the transition strategy towards sustainability. Their construct measuring sustainability practices of universities contains 1) community outreach programs; 2) green community centres; 3) partnerships with government, non-governmental organizations, and industry working toward sustainability; 4-5) written commitment to sustainability (university and department level); 6-7) sustainability audits on the surrounding community and on campus; 8) reuse of campus waste; 9) use of renewable and safe energy sources; 10) sustainable support services (e.g. recycling bins on campus, efficient public transport throughout the university); 11) sustainable campus building planning and 12) sustainable campus land-use.

Dagiliute, Liobikiene and Minelgaite (2018) were the first to investigate the differences in the perceived sustainability performance between the 'green' and the 'non-green' universities. They compared the students' attitudes towards sustainability in two Lithuanian universities. They did not find any significant differences in sustainability aspects in general, however, students of the green university sought more information about sustainability and were more often involved in sustainability activities. They also found that campus sustainability and environmental information have a significant impact on students' sustainable behaviour. In their study, they used a scale to measure perceptions made up of 16 items, grouped into four main constructs: 1) 'campus sustainability', 2) 'environmental information', 3) 'students' sustainability involvement', and 4) 'university's role in sustainable development. The item 'university's self-representation as a green university' was also involved in their construct. Their 17-item scale involves 1) environmental student organization(s); 2) use of public transport, bikes; 3) possibility to recycle waste at the university; 4) use one's own non-disposable cup; 5) availability of strategic documents and their implementation reports; 6) sustainability-related information during lectures;

7) university website on environmental objectives; 8) participation in environmental, social activities; 9) involvement in activities at the university; 10) energy and resource saving; 11) contribution to social well-being, tolerance; 12) environmental education; 13) cooperation with other national and foreign universities and businesses; 14) inclusion of sustainability aspects in study programmes; 15) sustainability research; 16) university's self-representation as a green university; and 17) declared environmental objectives. They found that students considered social aspects, i.e. social well-being, tolerance the most important attribute of the sustainable university. However, students considered environmental aspects, such as energy saving, environmental education, and actions less important.

Li, Gu and Liu (2018) established a new scoring system for campus sustainability in Australia. They suggest that sustainable campus performance indicators should be identified from the different perspectives of the economy, environment and society. In order to identify and prioritise the key sustainability indicators for university campuses, they proposed a new approach combining the qualitative scoring method and an analytical hierarchical process. After thorough literature review, they identified 54 indicators and quantified the weight coefficients for the criteria, sub-criteria and elements, and proposed a model that can be a flexible tool for university decision-makers.

*It is hypothesized that combining the most relevant items of the constructs developed by Nejati and Nejati (2013), Dagiliute, Liobikiene and Minelgaite (2018) and Li, Gu and Liu (2018), a new, reliable scale to measure perceived university sustainability, i.e. the sustainable university scale, can be developed (H2).*

Shuqin, et al. (2019) aimed to assess and compare the sustainability performance of different Chinese universities. The authors developed a campus sustainability evaluation system that is made up of the five main domains of campus sustainability, which are respectively: organization and management, energy and resource saving, friendly environment, campus culture, and social outreach. Their evaluation system included 14 mandatory indicators and 69 optional indicators. They found that the most problematic fields are organization management, resource saving and campus culture. For example, there are issues with green education, green research and green humanities as they are not so developed there. The assessment tool proposed by the authors can be used to guide the green campus revolution in China and could be adopted by the rest of the world.

Wakkee, et al. (2019) demonstrated how (entrepreneurial) universities can drive regional sustainable development in developing countries. They found that local campus leadership, a holistic teaching and research programme, and student involvement can have significant local effects.

### 1.2. Importance-Performance Analysis (IPA)

The importance-performance analysis (IPA) was developed by Martilla and James (1977). The original version of IPA defines consumer satisfaction as the function of two components that are respectively: the importance of an attribute of the product/service, and the perceived performance of the company on this attribute. The mean of importance and performance ratings of each attribute determines its position on the importance-performance matrix or grid, which is also often called the Cartesian diagram (Figure no. 1). The overall mean of the performance/importance ratings is used as a delimiter of high and low performance/importance (Yuvinatileng, Utomo and Latuperissa, 2013).

The 2x2 IPA matrix can be divided into four quadrants. Each quadrant requires a different approach and strategy (Wyród-Wróbel and Biesok, 2017):

• Quadrant 1: *Keep up the good work.* This is the best possible position for an attribute. This quadrant contains the competitive advantages and major strengths of a company. The organization must defend all of them to succeed. These are high importance/high performance items.

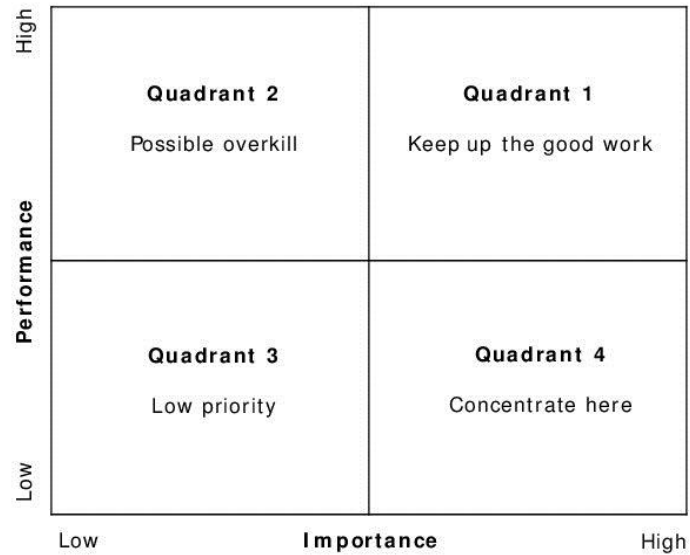

**Figure no. 1: The modified Importance Performance Matrix**
*Source: Kim, Jeon, Cho and Kim, 2018.*

• Quadrant 2: The territory of *Possible overkill*. Here low importance/high performance attributes, i.e. items of overperformance, can be found. Organizations should deploy business resources used here somewhere else (e.g. in Quadrant 1) or should increase the importance of those attributes that can be found here to turn them into competitive advantages.

• Quadrant 3: The area of *Low priority*. Low importance/low performance attributes can be seen here. Those are minor weaknesses that require no additional resources. Organizations are suggested to avoid investing in this quadrant.

• Quadrant 4: *Concentrate here*. High importance/low performance attributes can be found here. Those are the major weaknesses of an organization that require immediate corrective actions to increase consumer satisfaction and to avoid customer churn.

### 1.3. Stakeholder theory

The stakeholder theory originates from the 1980s. Freeman (1984) was the first to coin the phrase as an opposite to the shareholder theory or Friedman's doctrine, which suggests that a company's sole responsibility is to make money for its shareholders (Friedman, 1965).

According to the stakeholder theory, shareholders are only one of many stakeholders in a company, and an organization's key to market success is how it satisfies all the stakeholders, not only its shareholders (Freeman, 2010). The stakeholder theory says that the stakeholder ecosystem is made up of all parties that invested and involved in, or affected by, the company. Therefore, companies must pay special attention to their employees, vendors, suppliers, owners, community/neighbours, community groups, competitors, governmental bodies, oversight organizations and the local ecology (Freeman, 2010).

The stakeholder theory is intertwined with the domains of ethics and sustainability. Carroll and Buchholtz (2014) suggest that successful businesses in society adopt a stakeholder management approach. The stakeholder theory is solid ground for corporate social responsibility and business ethics inside the company (Kakabadse, Rozules and Davies, 2005).

The stakeholder ecosystem of a university comprises current, former (alumni) and potential students, parents, municipalities, academics, faculties, management (Rector, the Senate, Chancellor), administrative staff, governmental organisations, Academy of Sciences, research partners and companies. In higher education institutions, students and employees are always the major stakeholders in terms of their number. According to the stakeholder theory, universities are service providers to students and students are one of the most important stakeholders (Degtjarjova, Lapina, and Freidenfelds, 2018). The more satisfied students are, the more likely it is that the university could succeed, also in the field of sustainability. It is therefore assumed that IPA as a strategic tool should be used to maximize student satisfaction with the efforts that universities make towards sustainability.

## 2. Methodology

### 2.1. Methodology and research questions

Based on the literature review presented above, and in line with the main objectives of the research, this study aims to address the following research questions respectively:

- R1: What are the student expectations about university sustainability in Hungary? (student expectations)

- R2: To what extent are students satisfied with the sustainability performance of universities? (student satisfaction). H1 refers to this question.

- R3: Is combining sustainable university scale (SUS) with importance-performance analysis (IPA) a powerful strategic tool for university decision-makers to identify key areas of university sustainability?

- R4: What are the main components of the perceived university sustainability?

- R5: Is sustainable university scale (SUS) a reliable construct to measure students' perceptions of university sustainability? H2 refers to this question.

In line with the research questions, the following hypotheses were developed:

- H1: Students are not satisfied with the sustainability performance of the Hungarian higher education institutions.

• H2: Combining the most relevant items of the constructs developed by Nejati and Nejati (2013), Dagiliute, Liobikiene and Minelgaite (2018) and Li, Gu and Liu (2018), a new, reliable scale for measuring perceived university sustainability, i.e. the Sustainable University Scale (SUS), can be developed.

To answer the research questions, and to thoroughly investigate students' perceptions of the sustainable university, a questionnaire made up of 47 questions grouped into three sections were designed:

• Section 1: *Importance of the sustainable university scale (SUS) items*. It contains 21 statements measured on a five-point importance scale (1. not at all important … 5. very important). Respondents were asked to answer the following question: "How important are the followings to you regarding a sustainable university?". SUS items can be seen in Table no. 1.

• Section 2: *Perceived performance of the sustainable university scale (SUS) items*: The very same 21 statements as in Section 1, measured on a five-point rating scale (1 – very poor ... 5 – excellent), answering the question "How do you rate the sustainability performance of your university?".

• Section 3: *Demographic variables*. It contains 5 questions including gender, age, study level, branch of sciences and the university where they study (Table no. 2).

The sustainable university scale (SUS), which contains 21 items, is a construct developed by the authors. It is based on the domains of university sustainability discussed in the literature review. More specifically, in our construct we combined 9 items (item 4, 5, 7, 9, 10, 15, 17, 18 and 20) from Dagiliute, Liobikiene and Minelgaite (2018) with 9 items (item 1, 3, 6, 7, 8, 11, 13, 14 and 16) used by Nejati and Nejati (2013), with 3 items (item 9, 11 and 16) from Li, Gu and Liu (2018). It must be noted that four items are overlapping. They were found in not only one but two of the three reference studies (item 7, 9, 11 and 16). Moreover, we added four new items to SUS (item 2, 12, 19 and 21). The newly added items are 1) the awareness of the sustainability strategy of the university; 2) green location; 3) the inclusion of sustainability information into normal courses and 4) the integration of sustainability research results into the curricula. The sustainable university scale makes it possible that university decision-makers could gain deep insight into how students perceive their efforts towards sustainability.

Eight of 21 items were used without any modifications in its original form (referred as 'original'), nine items were modified to be unambiguous (referred to as 'revised'), and the four new items that we added are labelled as 'New' (Table no 1.).

**Table no. 1: The items of the sustainable university scale (SUS)**

|   | Sustainable university scale items | S* | Type |
|---|---|---|---|
| 1 | The university has a sustainability strategy | 2 | R |
| 2 | All the students, researchers, academic and non-academic staff are aware of the sustainability strategy of the university | 4 | N |
| 3 | Regular sustainability audits are performed on campus | 2 | O |
| 4 | Sustainability information is readily available on the university's website, newsletter, Neptun messages, etc. | 1 | R |

|     | **Sustainable university scale items** | **S*** | **Type** |
|-----|----------------------------------------|--------|----------|
| 5   | The university distinguishes itself as sustainable/green from other higher education institutions. | 1 | R |
| 6   | The university established environmentally and socially responsible purchasing practices | 2 | O |
| 7   | Separate waste collection is possible on campus, and the university encourages everyone to do so. | 1, 2 | R |
| 8   | The university uses renewable energy sources (e.g. solar panels). | 2 | O |
| 9   | The university saves water and energy (e.g. LED lighting) | 1, 3 | R |
| 10  | The university encourages use of public transport, bikes. | 1 | O |
| 11  | The university buildings are designed / converted in an energy efficient and sustainable way (e.g. windows, doors, insulation) | 2, 3 | R |
| 12  | The university buildings are located in a natural setting (quiet, green area with many trees where the air quality is excellent) | 4 | N |
| 13  | The university engages in community outreach programs that benefit the local environment. | 2 | O |
| 14  | The university has created partnerships with government, non-governmental organizations, and industry working toward sustainability. | 2 | O |
| 15  | The university has active environmental student organization(s) | 1 | O |
| 16  | There are many green actions, projects running / available at the university to support the achievement of sustainability goals | 2, 3 | R |
| 17  | The university offers a lot of study programmes related to sustainability. | 1 | R |
| 18  | The university offers a lot of subjects/courses related to sustainability. | 1 | R |
| 19  | There is also a lot of information about sustainability in normal courses | 4 | N |
| 20  | The university promotes sustainability research | 1 | O |
| 21  | Sustainability research results are integrated into the curricula | 4 | N |

*Notes: S\* (Source) = 1: Dagiliute, Liobikiene and Minelgaite (2018); 2: Nejati and Nejati (2013); 3: Li, Gu and Liu (2018); 4: New variables added by the authors.*
*Type = N: new, O: original, R: revised.*

An online survey, designed in Google Form, was conducted to collect data in October and November 2019. Current student status (ongoing studies) was the one and only eligibility criterion for students to participate in the study. Convenience sampling method was used. Students of nine Hungarian universities located in different regions of Hungary were asked to fill in the online questionnaire. The internal messaging systems of the universities were used to reach their students. Due to the low response rate, the sample size is 297.

SPSS 24 was used for data analysis (Babbie, Wagner and Zaino, 2019), and MS Excel for data visualisation (Walkenbach, 2016). Means were calculated to quantify the importance (R1) and performance (R2) of each item of the sustainable university scale. Importance-performance matrix was drawn to illustrate the position of SUS items to answer R3 (Kim et al., 2018.). To answer R4, Principle Component Analysis was run to understand patterns in SUS items (Jolliffe, 2011). The reliability of the sustainable university scale (SUS) was measured by Cronbach's alpha to answer R5 (DeVellis, 2017). Frequency tables of demographic variables were also calculated (Babbie, Wagner and Zaino, 2019).

### 2.2. The sample

Of the sample of 297 respondents, 61.3% was female, 38.7% male (Table no. 2). Mostly undergraduate students (77.1%) participated in this study, however some graduate students (16.8%) and doctoral students (6.1%) contributed to the survey. The majority of the respondents (54.2%) fell into the category 'aged 18-24'. Most of the students in the sample study social sciences (51.1%), engineering (23.9%) or humanities (13.87%). A significant part of them study in Miskolc (75.4%), the rest (24.6%) in other Hungarian universities. Therefore, this convenience sample is not representative, which is a limitation of this study.

Table no. 2: Distribution of demographic variables (N=297)

| Demographic variables | Values | Frequency | Percent |
|---|---|---|---|
| Gender | male | 115 | 38.7 |
|  | female | 182 | 61.3 |
| Study level | bachelor | 229 | 77.1 |
|  | master | 50 | 16.8 |
|  | PhD | 18 | 6.1 |
| Age | 18-24 | 161 | 54.2 |
|  | 25-31 | 60 | 20.2 |
|  | 32-38 | 29 | 9.8 |
|  | 39-45 | 27 | 9.1 |
|  | 46- | 20 | 6.7 |
| Branch of science | agricultural sciences | 2 | 0.7 |
|  | arts | 4 | 1.3 |
|  | engineering | 71 | 23.9 |
|  | humanities | 41 | 13.8 |
|  | medicine | 19 | 6.4 |
|  | natural sciences | 7 | 2.4 |
|  | social sciences | 153 | 51.5 |
| University | Corvinus University | 2 | 0.7 |
|  | Eszterhazy Uni. Eger | 14 | 4.7 |
|  | METU Budapest | 3 | 1.0 |
|  | National Uni. of Public Service | 2 | 0.7 |
|  | Szechenyi Uni. Gyor | 7 | 2.4 |
|  | University of Miskolc | 224 | 75.4 |
|  | University of Pannonia | 27 | 9.1 |
|  | University of Pecs | 15 | 5.1 |
|  | University of Szeged | 3 | 1.0 |

**3. Results and discussion**

This chapter is divided into five main sections, and each of them discusses the results related to one of the research questions.

**3.1. Importance of the sustainable university scale items (student expectations)**

In order to answer the first research question (R1), and to investigate student expectations about university sustainability in Hungary, the items of SUS were measured on a five-point importance scale. The lowest value (1) means the item is not at all important, whereas the highest value (5) indicates the item is very important. The importance of SUS items refers to the students' expectations regarding university sustainability. It expresses their opinion on what a university should do in order to be sustainable.

It was found that the opportunity for separate waste collection on campus and encouragement of this activity by the university is the most important attribute of university sustainability (4.54), whereas regular sustainability audits performed on campus is the least important for university students (3.51). They consider water and energy savings (e.g. the use of LEDs) as well as sustainable university buildings that are designed or converted in an energy efficient and sustainable way (e.g. windows, doors, insulation) extremely important (4.43).

If a university intends to be more sustainable, it must make efforts to provide the necessary infrastructure for sperate waste collection and promote this activity. The sustainable university should save water and energy and invest in sustainable, energy efficient buildings on campus. These findings are not fully consistent with those of Dagiliute, Liobikiene and Minelgaite (2018), who found recycling is less important for students.

It is also crucial for the students that university buildings must be located in a natural setting, e.g. in a quiet, green area with many trees where the air quality is excellent (4.39). Students, therefore, expect sustainable universities not only to be green, but to be located in a green environment. For the most important stakeholders, it is also essential that the sustainable university should use renewable energy sources, e.g. solar panels (4.35), it has a sustainability strategy (4.1) and promotes sustainability research (4.07).

It was also found that students think it important that the sustainable university carries out environmentally and socially responsible purchasing practices (4.0) and encourages the use of public transport, bikes (4.0). In a sustainable university, it is important that all the students, researchers, academic and non-academic staff should be aware of the sustainability strategy of the university (3.95) and sustainability information should be readily available on the university's website, newsletters, etc. (3.94), as well as the university should create partnerships with government, non-governmental organizations, and industry working toward sustainability (3.94). Green actions and projects (3.9) and community outreach programs (3.89) were found to be even less important.

The existence of environmental student organization(s) (3.76), the integration of sustainability research results into the curricula (3.72) as well as sustainability-focused positioning, when the university distinguishes itself as sustainable/green from other higher education institutions (3.71) are even less central for the students. There is only moderate demand for subjects/courses about sustainability (3.71) Students do not require that a lot of information about sustainability should be integrated into normal courses (3.61) or the university should offer a lot of study programmes related to sustainability issues (3.6). These results match those observed in earlier studies (Dagiliute, Liobikiene and Minelgaite, 2018). The overall mean of the importance items is 3.98. (Table no. 3.)

**3.2. Perceived performance of the sustainable university items (student satisfaction)**

In order to answer the second research question (R2), and to find out to what extent students are satisfied with the performance of the Hungarian universities towards sustainability, students were asked to rate the performance of the universities on a five-point rating scale. The lowest score (1) indicates *very poor* rating (dissatisfaction), whereas the highest score (5) means *excellent* rating (very high satisfaction). The rating scores of the sustainable university scale items refer to how students are satisfied with the sustainability performance of the university where they study. It expresses their opinion on how sustainable the university is perceived regarding each attribute (item) of the sustainable university scale. It allows decision-makers to get more insight into how their efforts towards sustainability are seen by their students, their most important stakeholders.

As far as the perceived sustainability performance of the Hungarian universities is concerned, their overall performance rating is only 3.23, which means that that students are not satisfied with it and consider Hungarian universities unsustainable (Table no. 3). These results provide support for the first hypothesis (H1), therefore it has been accepted.

Students are most satisfied with the location of the university buildings, the rating of which is very good (4.17). It suggests that Hungarian universities have preferred locations that are mostly situated in quiet, green areas with many trees where the air quality is excellent. This could be a strength they capitalise on. Student are also satisfied with the separate waste collection opportunities on campus (3.7), community outreach programs benefiting the local environment (3.5) and the promotion of sustainability research (3.47). Students are somewhat satisfied with the efforts made towards sustainability strategy (3.35), partnerships with government, non-governmental organizations, and industry working towards sustainability (3.34), as well as sustainable university buildings (3.33), water and energy savings in the university (3.29) and the use of public transport and bikes (3.27). However, students are not really satisfied with how much information about sustainability is integrated into normal courses (3.14) and the mostly unsustainable purchasing practices of universities (3.13). They are not convinced by the green actions/projects (3.12) and the integration of sustainability research results into the curricula (3.11).

Moreover, students think only limited information on sustainability is available for them on the website or in the newsletters of the universities (3.08). This is a serious problem as the lack of information is usually one of the greatest barriers towards sustainability (Avila et al, 2017). Also, students think that they and other important stakeholders (researchers, academic and non-academic staff) are not aware of the sustainability strategy of the university (3.06), however their participation would be essential in the implementation.

Students do not think that universities position themselves as sustainable/green (3.03) or use solar panels or other renewable energy sources (3.02). They are not content with the number of subjects/courses about sustainability (3.0), green/environmental student organizations (2.99) and the number of study programmes related to sustainability (2.93). Students were found to be the least satisfied with the sustainability audits on campus (2.82).

### 3.3. Importance-performance analysis (IPA)

In this section, in line with research question 3 (R3), it is discussed whether combining the sustainable university scale (SUS) with importance-performance analysis (IPA) could be a useful strategic tool for university decision-makers to identify key areas of university sustainability. In order to determine the position of each item of the sustainable university scale in the quadrants of the importance-performance matrix, deviations of the means from the overall mean of importance (Δ IMP) and performance (Δ PER) were calculated. Table no. 3 shows the results and the position of each item in the quadrants of IPA.

Seven attributes of the sustainable university scale including location, separate waste collection, strategy, energy and water savings, public transport, research and sustainable buildings fall into the 'Keep up the good work" quadrant (Q1), which contains the competitive advantages (strengths) of the Hungarian universities with regard to sustainability. It is suggested that universities should use all of them in communication campaigns targeted at students who are concerned about sustainability.

**Table no. 3: Importance and performance of the sustainable university scale items**

|    |                                   | IMP means | PER means | Δ IMP | Δ PER | Quad-rant |
|----|-----------------------------------|-----------|-----------|-------|-------|-----------|
| 1  | Sustainability strategy           | 4.10      | 3.35      | 0.12  | 0.12  | Q1        |
| 2  | Awareness of the sust. strategy   | 3.95      | 3.06      | -0.03 | -0.17 | Q3        |
| 3  | Sustainability audits             | 3.51      | 2.82      | -0.47 | -0.41 | Q3        |
| 4  | Sustainability information        | 3.94      | 3.08      | -0.04 | -0.15 | Q3        |
| 5  | Green positioning                 | 3.71      | 3.03      | -0.27 | -0.20 | Q3        |
| 6  | Green purchasing                  | 4.00      | 3.13      | 0.02  | -0.10 | Q4        |
| 7  | Separate waste collection         | 4.54      | 3.70      | 0.56  | 0.47  | Q1        |
| 8  | Renewable energy sources          | 4.35      | 3.02      | 0.37  | -0.21 | Q4        |
| 9  | Water and energy savings          | 4.43      | 3.29      | 0.45  | 0.06  | Q1        |
| 10 | Public transport, bikes           | 4.00      | 3.27      | 0.02  | 0.04  | Q1        |
| 11 | Sustainable buildings             | 4.43      | 3.33      | 0.45  | 0.10  | Q1        |
| 12 | Green location                    | 4.39      | 4.17      | 0.41  | 0.94  | Q1        |
| 13 | Community outreach programs       | 3.89      | 3.50      | -0.09 | 0.27  | Q2        |
| 14 | Sustainability partnerships       | 3.94      | 3.34      | -0.04 | 0.11  | Q2        |
| 15 | Green student organization(s)     | 3.76      | 2.99      | -0.22 | -0.24 | Q3        |
| 16 | Green actions, projects           | 3.90      | 3.12      | -0.08 | -0.11 | Q3        |
| 17 | Green study programmes            | 3.60      | 2.93      | -0.38 | -0.30 | Q3        |
| 18 | Green subjects/courses            | 3.71      | 3.00      | -0.27 | -0.23 | Q3        |
| 19 | Greening normal courses           | 3.61      | 3.14      | -0.37 | -0.09 | Q3        |
| 20 | Sustainability research           | 4.07      | 3.47      | 0.09  | 0.24  | Q1        |
| 21 | Sustainability research integration | 3.72    | 3.11      | -0.26 | -0.12 | Q3        |
|    | Total                             | 3.98      | 3.23      |       |       |           |

*Notes: IMP: importance; PER: performance; Quadrants: (1) Keep up the good work (2) Possible overkill (3) Low priority (4) Concentrate here*

Campus location is found to be the biggest strength. The favourable location is very important for the students. They require that university buildings should be situated in a quiet, green environment, and for most of them, this expectation is fully met. Separate waste collection, which is the most important aspect of the sustainable university from the students' perspective, is also a major strength as students are quite satisfied with it. Hungarian universities must communicate that they provide the infrastructure for separate waste collection and promote this activity.

Based on our findings, it is advisable for universities to emphasize that their students are satisfied with their efforts towards energy and water savings and appreciate their endeavours to increase energy efficiency on campus. Also, students are content with how sustainable the design of the university buildings is. It can also be suggested that Hungarian higher education institutions should communicate that they promote sustainability research, encourages the use of public transport, bikes and they have a written sustainability strategy.

Two items can be found in Q2, which is *the possible overkill* quadrant. It contains items that are not important for the students, however they, the most important stakeholders of the universities are satisfied with it (performance ratings are better than the overall average). The performance of universities concerning community outreach programs, partnership with governmental, non-governmental organizations, and industry is better than required. In this case, it is suggested that universities should make a communication campaign to increase the importance of their community outreach programs and sustainability partnership to turn those activities into competitive advantages.

Ten items – nearly the half of the sustainable university scale items – can be found in Q3, which represents "Low priority" attributes having low importance and low perceived performance. The items that fall into this quadrant are respectively: 1) awareness of the sustainability strategy; 2) regular sustainability audits; 3) information regarding sustainability (website, newsletters, etc.); 4) sustainability-focused positioning of the universities; 5) active green student organizations; 6) sustainability-related projects/actions; 7) study programs related to sustainability; 8) subjects/courses related to sustainability; 9) integration of sustainability into normal/traditional courses; and finally, 10) integration of sustainability research results into the curricula. Hungarian universities are strongly advised to avoid any investments in those activities.

Last but not at least, two sustainable university items can be found in Q4. This is the "Concentrate here" quadrant representing attributes that universities should immediately improve to achieve higher student satisfaction with regard to their attempts to be more sustainable. The items listed here have high importance and low perceived performance suggesting that students are really dissatisfied with them in spite of the fact that those items are really important for them. On the one hand, they do not believe that universities have environmentally and socially responsible purchasing practices, on the other hand they are disappointed with the use renewable energy sources (e.g. solar panels) on campus. It is therefore suggested that universities should concentrate more on green/socially responsible procurement and should increase the use of renewable energy sources to make students who are concerned about sustainability more satisfied. Universities should consider more the sustainability performance of their suppliers. They should be greening their tenders, prefer local suppliers, and install more solar panels, etc. (Figure no. 2).

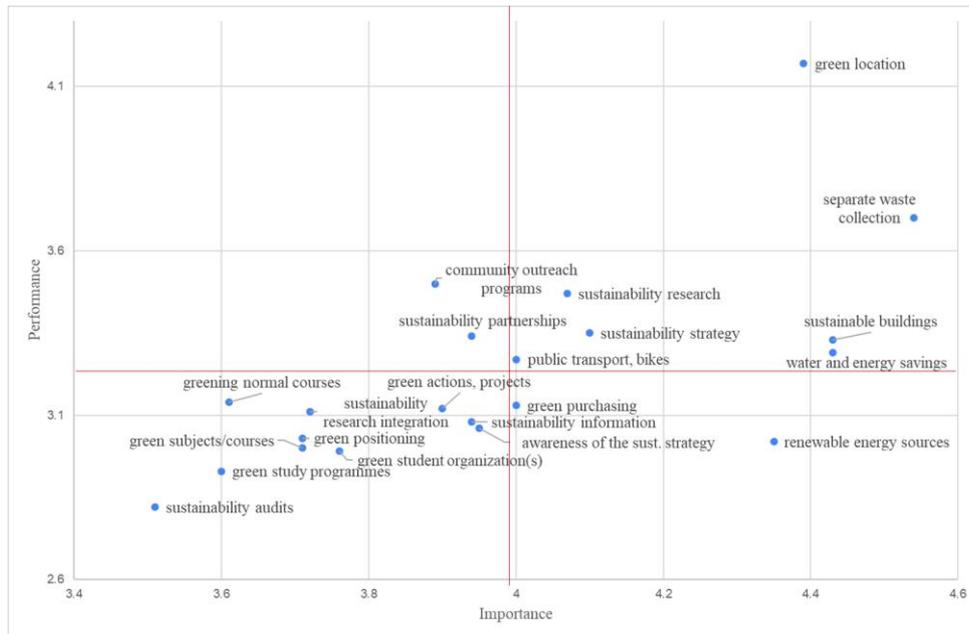

**Figure no. 2: Importance-performance of the sustainable university scale items**

As no research has been found that surveyed the perceived importance and performance of the attributes of university sustainability, it is therefore not possible to compare the results discussed here to the findings of previous works. However, this study fills this gap in the literature and propose a new methodology to investigate the attributes of university sustainability. As an answer to R3, it can be concluded that importance performance analysis (IPA) is a strong strategic tool for university decision-makers to identify key areas of university sustainability when combined with the sustainable university scale (SUS). Using the results of IPA, universities could implement corrective actions to make students as stakeholders more satisfied with their efforts to be more sustainable.

**3.4. Factor analysis of the sustainable university scale items**

In order to investigate patterns in perceived university sustainability, and answer R4, factor analysis was used. The dataset of the *importance of SUS items* were analysed as it refers to the students' expectation. The very high Kaiser-Meyer-Olkin Measure of Sampling Adequacy value (KMO=0.938) indicates that a factor analysis is a useful method with our data. The Bartlett's Test of Sphericity (Approx. Chi-Square = 4400.484; df = 210; Sig.= 0.000) also reconfirms it (Jolliffe, 2011).

The extraction communalities are acceptable, although the lower values of *Green Location* and *Green Positioning* show that they don't fit as well as the others. Only three factors in the initial solution have eigenvalues greater than 1. Together, they account for almost 65% of the variability in the original variables (Table no. 4)**.** This suggests that three latent influences are associated with sustainable university perceptions, but there remains room for a lot of unexplained variation (Babbie, Wagner and Zaino, 2019). The scree plot also confirmed the choice of three components.

**Table no. 4: Total variance explained**

| Compo-nent | Initial Eigenvalues | | | Extraction Sums of Squared Loadings | | | Rotation Sums of Squared Loadings | | |
|---|---|---|---|---|---|---|---|---|---|
| | Total | % of Variance | Cumulative % | Total | % of Variance | Cumulative % | Total | % of Variance | Cumulative % |
| 1 | 10.648 | 50.706 | 50.706 | 10.648 | 50.706 | 50.706 | 5.782 | 27.535 | 27.535 |
| 2 | 1.650 | 7.856 | 58.563 | 1.650 | 7.856 | 58.563 | 3.937 | 18.746 | 46.281 |
| 3 | 1.333 | 6.346 | 64.909 | 1.333 | 6.346 | 64.909 | 3.912 | 18.628 | 64.909 |
| 4 | .916 | 4.360 | 69.269 | | | | | | |
| 21 | .139 | .663 | 100.000 | | | | | | |

*Notes: Extraction Method: Principal Component Analysis.*

To rotate the factor components, Varimax rotation with Kaiser normalization was used. The first rotated factor component is most highly correlated with *community outreach programs, sustainability partnerships, green study programmes, green subjects/courses, greening normal* courses*, sustainability research and sustainability research integration* items (Table no. 5). These variables are not particularly correlated with the other two factor components, and each of them refers to actions towards meeting sustainability objectives, or related to education or research, it is therefore the first component called as *Sustainable Actions, Education & Research (SAER)*.

**Table no. 5: Rotated component matrix**

| Items | 1. Sust. Actions, Education & Research | 2. Sust. Operation/ Infrastructure | 3. Sust. Strategy | Type |
|---|---|---|---|---|
| Sustainability strategy | 0.20 | 0.23 | **0.76** | ST1 |
| Awareness of the sust. strategy | 0.24 | 0.24 | **0.80** | ST2 |
| Sustainability audits | 0.39 | 0.06 | **0.73** | ST3 |
| Sustainability information | 0.25 | 0.32 | **0.74** | ST4 |
| Green positioning | 0.31 | 0.29 | **0.48** | ST5 |
| Green purchasing | **0.51** | 0.37 | 0.44 | PU1 |
| Separate waste collection | 0.15 | **0.78** | 0.24 | WE1 |
| Renewable energy sources | 0.26 | **0.75** | 0.30 | WE2 |
| Water and energy savings | 0.13 | **0.82** | 0.31 | WE3 |
| Public transport, bikes | 0.50 | **0.56** | -0.01 | WE4 |
| Sustainable buildings | 0.28 | **0.73** | 0.26 | WE5 |
| Green location | 0.39 | **0.49** | 0.08 | LO1 |
| Community outreach programs | **0.60** | 0.31 | 0.33 | SA1 |
| Sustainability partnerships | **0.68** | 0.25 | 0.33 | SA2 |
| Green student organization(s) | **0.57** | 0.33 | 0.41 | SA3 |
| Green actions, projects | **0.63** | 0.28 | 0.45 | SA4 |
| Green study programmes | **0.83** | 0.16 | 0.19 | SER1 |
| Green subjects/courses | **0.82** | 0.17 | 0.21 | SER2 |
| Greening normal courses | **0.71** | 0.16 | 0.27 | SER3 |
| Sustainability research | **0.71** | 0.34 | 0.30 | SER4 |
| Sustainability research | **0.78** | 0.22 | 0.26 | SER5 |

| Items | 1. Sust. Actions, Education & Research | 2. Sust. Operation/ Infrastructure | 3. Sust. Strategy | Type |
|---|---|---|---|---|
| integration | | | | |

*Notes: (1) Extraction Method: Principal Component Analysis. Rotation Method: Varimax with Kaiser Normalization. Rotation converged in 6 iterations. (2) ST: Strategy, commitment & monitoring; PU: Purchasing; WE: Waste & energy; LO: Location; SA: Sustainability actions; SER: Sustainable education & research*

The second factor component, which is called *Sustainable Operation/Infrastructure,* are made up of s*eparate waste collection, renewable energy sources, water and energy savings and sustainable buildings.* All those items are related to the domain of waste and energy. The third component, *Sustainable Strategy,* has been named after the items that correlated with it the most. All of them are related to the sustainability strategy including the written *sustainability strategy*, and its *awareness,* regular *sustainability audits* and *sustainability information*. Because of their moderately large correlations with the first and the third components, *green student organizations and green action/projects* bridges *Sustainable Actions, Education & Research* and *Sustainable Strategy*. *Public transport, bikes* and *green location* variables bridge the first and the second components, whereas *green positioning* and *green purchasing* are highly correlated with all the three factor components.

These results suggest that students form expectations about the three main domains of university sustainability: 1) sustainable strategy, 2) sustainable operations/infrastructure, and 3) sustainable actions/education/research. These are the main topics of the university sustainability in the mind of the most important stakeholder. These findings are not in line with those of previous studies (Nejati and Nejati, 2013; Dagiliute, Liobikiene and Minelgaite, 2018). In both of the earlier studies, the number of factor components was higher, and the structure of the component was different from our results.

It is proposed that universities should deal with all the three components separately, and it would be beneficial for them to assign managers in charge to each domain to fully meet student expectations.

### 3.5. Reliability of the sustainable university scale

In order to test H2 and to investigate whether all the 21 items of the sustainable university scale reliably measure the same latent variable, a Cronbach's alpha was run on both SUS importance and SUS performance datasets.

In the reliability statistics table of SUS importance, Cronbach's alpha was 0.95, which indicates a very high level of internal consistency for our scale with this specific sample (DeVellis, 2017.). The "Cronbach's Alpha "If Item Deleted" column showed that removal of any item would result in a lower Cronbach's alpha, so no items were removed from the 21 item-scale. In the reliability statistics table of SUS performance dataset, Cronbach's alpha was even higher (0.985), which indicates an even higher level of internal consistency. Here also no items were removed as the "If Item Deleted" column showed that removal of any item would result in a lower Cronbach's alpha.

Also, a reliability analysis was run in order to ensure internal consistency of the identified constructs after the principle component analysis. The high Cronbach's alpha values confirmed the reliability of the constructs (α Sustainable Strategy = 0.850, no. of items = 5;

α Sustainable Operation & Infrastructure = 0.861, no. of items = 6; and α Sustainable Actions, Education & Research= 0.938, no. of items =10).

All these findings support H2. It is therefore accepted that the sustainable university scale (SUS) is a reliable construct to measure perceived university sustainability.

**Conclusions**

The stakeholder theory suggests that organizations should fully meet stakeholders' expectations to be successful (Freeman, 2010). Students are one of the biggest and most important stakeholders of universities (Degtjarjova, Lapina and Freidenfelds, 2018), and could have a significant impact on the environment (Emanuel and Adams, 2011). Nowadays, the public demand for more sustainable universities is growing (Md Shahbudin, et al., 2011.). More and more students want to study about sustainability, expect the integration of sustainability research into curricula and prefer universities that make efforts to operate in a more sustainable manner (Dagiliute, Liobikiene and Minelgaite, 2018). University decision-makers (Rector, Chancellor, Deans and the Senate) should consider sustainability issues to a greater extent when developing organizational strategy. This study extends the knowledge of the above decision-makers regarding students' perception of university sustainability in many aspects.

The current study found that separate waste collection on campus is the most important student expectation about sustainability. However, it is not in line with the result of previous studies. Dagiliute, Liobikiene and Minelgaite (2018) found recycling less important for students. Nonetheless, our findings are consistent with those of other studies suggesting that students expect water and energy savings and energy efficient, sustainable university buildings in a sustainable university. Also, it is important for the students that the buildings should be located in a green environment. Universities are therefore advised to promote separate waste collection, save water and energy, and maintain sustainable, energy efficient buildings that are situated in green parks (R1).

In the current study, the low value of general satisfaction with the performance of universities towards sustainability (3.23) confirmed H1 and suggests that students are not satisfied with it and consider Hungarian universities rather unsustainable. Students' perceptions of university sustainability are in line with the weak positions of the Hungarian higher education institution in green rankings (Greenmetric, 2019). Our findings show that students are most satisfied with the location of the university buildings, which suggests that Hungarian universities have preferred locations. Also, students are content with the opportunity to collect waste separately on campus, the community outreach programs that universities offer, and the promotion of research on sustainability (R2).

The findings of this research confirmed H3. By combining the importance-performance analysis (IPA) with the sustainable university scale (SUS), a simple but powerful strategic managerial tool can be developed. It could be widely used by university decision-makers to investigate the key areas of university sustainability. IPA helps to identify competitive advantages and major weaknesses in the domains of sustainability and make it possible for decision-makers to implement corrective actions to make students as stakeholders more satisfied with the university's efforts to address sustainability. Two major weaknesses were found in our study. Hungarian universities perform poorly in sustainable purchasing and use less renewable energy (e.g. solar panels) on campus than it is expected by their students. It is therefore suggested that universities should immediately make both their energy use and purchasing process more sustainable. On the other hand, it was also found that campus location and separate waste collection are the major competitive advantages. It is suggested that the major strengths are used in the marketing campaigns of universities to make their green positioning more effective and to build the sustainable university brand image. Strategy, energy and water savings, public transport, sustainable buildings and research are also strengths of the Hungarian universities that should be communicated (R3).

The three main domains of university sustainability were also identified. These are the strategy towards sustainability, actions to promote sustainability including education and research, and the sustainable infrastructure/operations. This is a unique structure and different from those presented in earlier studies (Nejati and Nejati, 2013; Dagiliute, Liobikiene and Minelgaite, 2018), which suggests that Hungarian universities should use a nation-specific approach to university sustainability. Future studies on this topic are therefore recommended to investigate it in different cultural and national contexts (R4).

The sustainable university scale (SUS) was found to be a reliable construct to measure perceived university sustainability (H2 accepted). The adaptation of this construct is therefore proposed to both researchers and university decision-makers to investigate how students do perceive the efforts that universities make towards sustainability. Combined with IPA, it could be a powerful benchmarking tool, which is an important practical implication (R5).

Further research should be done to compare the perceived university sustainability of green and non-green universities in different cultural settings.